# The HARPS search for southern extra-solar planets

## I. HD 330075 b: a new "hot Jupiter"

F. Pepe[1], M. Mayor[1], D. Queloz[1], W. Benz[2], X. Bonfils[1,3], F. Bouchy[4], G. Lo Curto[5], C. Lovis[1], D. Mégevand[1], C. Moutou[4], D. Naef[5,1], G. Rupprecht[6], N. C. Santos[7,1], J.-P. Sivan[4], D. Sosnowska[1], and S. Udry[1]

[1] Observatoire de Genève, 51 ch. des Maillettes, CH–1290 Sauverny, Switzerland
[2] Physikalisches Institut Universität Bern, Sidlerstrasse 5, CH3012 Bern, Switzerland
[3] Laboratoire d'Astrophysique, Observatoire de Grenoble, B.P. 53, 414 rue de la Piscine, F–38041 Grenoble CEDEX 9, France
[4] Laboratoire d'Astrophysique de Marseille, Traverse du Siphon BP8, F–13376 Marseille CEDEX 12, France
[5] ESO La Silla Observatory, Alonso de Cordova 3107, Vitacura Casilla 19001, CL–Santiago 19, Chile
[6] ESO Head Quarter, Karl-Schwarzschild-Str. 2, D–85748 Garching b. München, Germany
[7] Observatório Astronómico de Lisboa, Tapada da Ajuda, P–1349-018 Lisboa, Portugal

**Abstract.** We report on the first extra-solar planet discovered with the brand new HARPS instrument. The planet is a typical "hot Jupiter" with $m_2 \sin i = 0.62\,M_{Jup}$ and an orbital period of 3.39 days, but from the photometric follow-up of its parent star HD 330075 we can exclude the presence of a transit. The induced radial-velocity variations exceed $100\,\text{m s}^{-1}$ in semi-amplitude and are easily detected by state-of-the-art spectro-velocimeters. Nevertheless, the faint magnitude of the parent star ($V = 9.36$) benefits from the efficient instrument: With HARPS less than 10 observing nights and 3 hours of *total* integration time were needed to discover the planet and characterize its orbit. The orbital parameters determined from the observations made during the first HARPS run in July 2003 have been confirmed by **7** additional observations carried out in February 2004. The bisector analysis and a photometric follow-up give no hint for activity-induced radial-velocity variations, indicating that the velocity curve is best explained by the presence of a low-mass companion to the star. In this paper we present a set of 21 measurements of excellent quality with weighted *rms* as low as $2.0\,\text{m s}^{-1}$. These measurements lead to a well defined orbit and consequently to the precise orbital parameters determination of the extra-solar planet HD 330075 b.

## 1. Introduction

The majority of the known extra-solar planets in the southern hemisphere have been discovered using our CORALIE instrument on the Swiss-Euler-1.2m-Telescope at the ESO-La Silla Observatory, Chile (see Mayor et al. 2004, and references therein)[1]. Up to now, the small telescope size could be compensated by adopting efficient observation strategies. In particular we have been using the simultaneous ThAr-lamp technique – which is probably the most efficient calibration technique (Bouchy et al. 2001) – allowing us to compete with other instruments on much larger telescopes. Precisions down to $3\,\text{m s}^{-1}$ have been attained with CORALIE on bright stars (Queloz et al. 2001b). Nevertheless, the photon noise remained very often the limiting factor in our measurements. Only with a new instrument on a bigger telescope we would be able to beat this limitation.

The installation of HARPS[2] on the ESO 3.6-m telescope during 2003 provided us with the suited access to even more precise and efficient radial-velocity measurements. In fact, the HARPS instrument was manufactured for the European Southern Observatory (ESO) by the HARPS Consortium[3] in exchange for Guaranteed Time Observation (GTO) on the ESO 3.6-m telescope. The first GTO period took place in July 2003 providing immediately 480 high-quality and high-resolution stellar spectra in 9.5 nights. The great efficiency of HARPS is emphasized by the automatic and real-time data-reduction

---

*Send offprint requests to*:
F. Pepe, e-mail: `Francesco.Pepe@obs.unige.ch`

[1] We refer also to Endl et al. (2003); Hatzes et al. (2003); Perrier et al. (2003); Tinney et al. (2002); Endl et al. (2002); Vogt et al. (2002); Fischer et al. (2001); Udry et al. (2000); Noyes & et al. (1997) for the description of the various ongoing extra-solar planet surveys

[2] High-Accuracy Radial-velocity Planet Searcher

[3] The Consortium is composed of the Geneva Observatory (leading institute), the Observatoire de Haute-Provence, ESO, the Physikalisches Institut der Universität Bern, and the Service d'Aéronomie du CNRS

pipeline which delivered, already by the first GTO run, reduced, wavelength-calibrated spectra and, of course, the precise radial velocity of the observed star. This allowed us to focus immediately on the analysis of the data acquired during the first nights. Among the large number of measurements we were able to isolate a bunch of variable objects potentially hosting short-period extra-solar planets. These objects were then selected for an intensive follow-up during the remaining nights. As a result a Jupiter-mass companion to HD 330075 was discovered and its orbit characterized.

The discovery of HD 330075 b was announced in December 2003 (Mayor et al. 2003) after careful analysis of the line bisector and a photometric follow-up of the parent star, in order to exclude the possibility of activity-induced radial-velocity variations. A Jupiter-mass companion remained the best explanation for the observed radial-velocity curve of HD 330075. Nevertheless we wanted to wait for the new observational season to check for consistency between the data sets. Indeed, the new data plotted very well on the curve determined from the first set of measurements taken 6 months before.

## 2. Observing with the HARPS spectrograph

HARPS is ESO's "High-Accuracy Radial-velocity Planet Searcher" (Mayor et al. 2003; Pepe et al. 2002). It is installed on the 3.6-m telescope at the La Silla Observatory and available to the astronomical community since October 1st, 2003. The instrument is optimized for precise radial-velocity measurements in general, and for the detection of exoplanets by means of the Doppler technique in particular.

The HARPS instrument is a fiber-fed high-resolution echelle spectrograph. Its optical design is similar to that of the UVES spectrograph on the VLT and therefore based on a concept which has already demonstrated its high optical efficiency. The covered spectral domain is $3800 < \lambda < 6900$ and the spectral resolution is of $R = 115'000$ for the fiber diameter of 1 arcsec. The signal-to-noise ratio obtained in a 1-minute integration on a 6th magnitude star is $SNR = 100$ per spectral bin ($\lambda = 550$ nm, $\Delta\lambda = 0.00147$ nm). The conversion from signal-to-noise ratio to photon-noise precision depends only on the stellar spectral type and rotational velocity. For example we obtain at $SNR = 100$ a precision of $1$ m s$^{-1}$ for a non-rotating G-dwarf. It is worth to mention that the relation is strictly inversely proportional, even for low signal-to-noise values aproaching unity.

But the real strength of HARPS compared to UVES and other high-resolution spectrographs is its extra-ordinary instrumental stability. Great efforts have been made to stabilize the instrumental profile (IP) and the physical position of the spectrum on the CCD. This goal is achieved by several design choices: a) The instrument is installed in the Coudé Room of the telescope building and does not move. b) It is fed by fibers which guarantee excellent input-beam stability compared to slit spectrographs. This property is further improved by introducing a double image scrambler into the light path. c) The instrument is operated in vacuum in order to avoid drifts of the spectrum on the CCD due to changes in atmospheric pressure. d) The temperature of the spectrograph is kept stable to better than 0.01 K over a year. e) Both spectral resolution and line sampling (3.5 px per FWHM) have been chosen as high as possible in order to minimize the impact of possible IP-variations on the radial-velocity measurement. The stability thus obtained is remarkable: During a single observing night the instrumental drift remains always well within $1$ m s$^{-1}$ in absolute terms, i.e. without adopting any kind of calibration or an a-posteriori correction. The stability has been verified by using the simultaneous ThAr technique (Baranne et al. 1996) which is able to track instrumental drifts with a precision of better than $0.09$ m s$^{-1}$ (Mayor et al. 2003).

The internal stability of the instrument as well as the input illumination stability are of fundamental importance when aiming at highest precision. It has been shown by Butler et al. (2004) that guiding errors or spectrograph temperature drifts can lead to radial-velocity jumps and drifts of up to $10$ m s$^{-1}$. In the case of HARPS these kind of problems have been removed "at the source" by means of the solutions described above. In particular the guiding problems are considerably reduced by the scrambling properties of the fibers. Thus, instrument drifts and IP variations are almost completely avoided. Possible residual drifts are eventually removed by applying the simultaneous ThAr technique described in Baranne et al. (1996).

The short-term precision performance of HARPS has been characterized on the sky by intensive asteroseismology observations carried out during the instrument commissioning phase (Mayor et al. 2003). The observations proved that on short time-scales (one night) the precision is better than $0.2$ m s$^{-1}$ *rms* and mainly limited by photon noise and, in particular, by the intrinsic stability of the star itself. Besides tracking the instrument drifts the ThAr-lamp does also ensure the long-term precision of the instrument. Daily calibrations allow us to remove long-term trends, if any. It had been shown already with CORALIE that a precision of $3$ m s$^{-1}$ *rms* over more than 4 years can be attained. The first months of operation with HARPS indicate that even better long-term precision is within reach.

For efficient observation an extra-ordinary spectrograph alone is not sufficient. Nowadays, a high-performance data-reduction must be part of the observing facility. The data-reduction pipeline of HARPS offers a complete treatment of the data. The extracted and wavelength calibrated spectra as well as the precise stellar radial velocity are delivered to the observer about 30 seconds after the end of the exposures with final accuracy.

## 3. A hot Jupiter around HD 330075

### 3.1. Stellar characteristics of HD 330075

The HIPPARCOS catalogue (ESA 1997) classifies HD 330075 as a G5 dwarf. Its magnitude is $V = 9.36$ while the color index is given as $B - V = 0.935$. The precise astrometric parallax was measured to $\pi = 19.92$ mas corresponding to a distance of about 50 pc from the Sun. The derived absolute magnitude is $M_V = 5.86$. Atmospheric parameters such as effective temperature $T_{\text{eff}} = 5017$ K, surface gravity $\log g = 4.22$, as well as [Fe/H] $= +0.08$ have been derived in a detailed LTE spectroscopic analysis using the method presented in Santos

**Table 1.** Observed and inferred stellar parameters for HD 330075

| Parameter | | HD 330075 |
|---|---|---|
| Spectral Type | | K1 |
| V | | 9.36 |
| $B - V$ | | 0.935 |
| $\pi$ | [mas] | $19.92 \pm 1.5$ |
| Distance | [pc] | $50.20 \pm 4$ |
| $M_V$ | | 5.86 |
| $L/L_\odot$ | | 0.47 |
| $[Fe/H]$ | | $0.08 \pm 0.06$ |
| $M/M_\odot$ | | $0.7 \pm 0.1$ |
| $T_{\rm eff}$ | [K] | $5017 \pm 53$ |
| $\log g$ | [cgs] | $4.22 \pm 0.11$ |
| $v \sin i$ | [km s$^{-1}$] | $0.7 \pm 0.2$ |
| $\log(R'_{\rm HK})$ | | $-5.03$ |
| $P_{\rm rot}(R'v_{\rm HK})$ | [days] | 48 |
| age$(R'_{\rm HK})$ | [Gyr] | 6.2 |

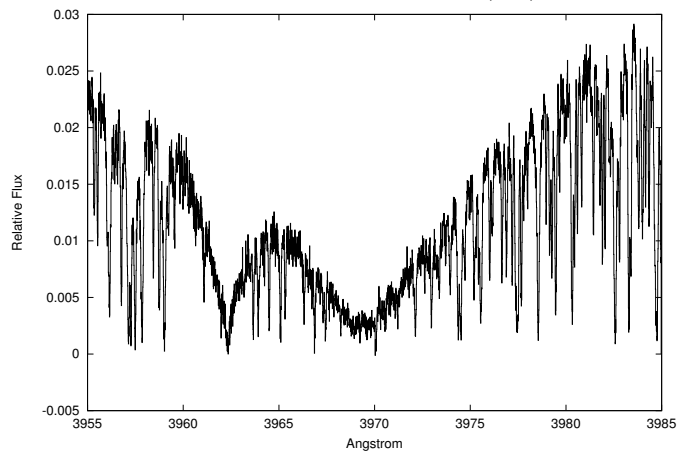

**Fig. 1.** $\lambda 3968.5$ Å Ca II H absorption line region of a single HARPS spectrum of HD 330075. The absence of an emission peak in the center of the absorption line is an indication for rather low chromospheric activity

et al. (2004). The same authors give a formula to derive the HIPPARCOS $\log_{HIP} g = 4.37$, which is in excellent agreement with the spectroscopic value. The effective temperature and the $B - V$, however, are in contrast to the spectral type given in the HIPPARCOS catalogue. Considering the almost solar metallicity we would rather expect HD 330075 to be a K1 to K2 star. $\log g$ is however below the value expected for this spectral type, and we deduce therefore that the star is probably slightly evolved. This is as well consistent with the higher measured luminosity $L = 0.47 L_\odot$ compared to a K1 dwarf (typically $0.37 L_\odot$). From the absolute magnitude and the $B-V$ we finally estimate a mass of about $M = 0.7 M_\odot$. The stellar parameters are summarized in Table 1.

Fig. 1 shows the Ca II H absorption line region of a single HARPS spectrum at $\lambda 3968.5$ Å. The emission flux in the core of the Ca II H line corrected for the photospheric flux provides us with the chromospheric activity indicator $\log(R'_{\rm HK}) = -5.03$. This value is typical for stars with very low chromospheric activity level (Henry et al. 1996). Using the calibration given by Donahue (1993) and quoted in Henry et al. (1996) we compute for this star an age of approximately 6 Gyrs, while the rotational period resulting from the calibration given by Noyes et al. (1984) is of 48 days. We have estimated the projected rotational velocity of the star to be $v \sin i = 0.7$ km s$^{-1}$ by taking one additional measurement with the CORALIE instrument, for which the cross-correlation function is well characterized. Because of the small rotational velocity and the low activity level of the star we do not expect any significant velocity jitter. From the CORALIE measurement and following (Santos et al. 2002) we also get an independent determination of the metallicity [Fe/H] = +0.06, which is almost identical to the value derived from the HARPS spectrum.

### 3.2. HARPS orbital solution for HD 330075 b

The precise radial-velocity data of HD 330075 have been obtained by our group during two observing runs in August 2003 and February 2004, respectively[4]. The campaigns delivered a total of 21 data points. In exposures lasting between 6 and 15 minutes we obtained individual photon-noise errors of 1 m s$^{-1}$ under good weather conditions and 4 m s$^{-1}$ in the worst case. These photon noise values represent the fit uncertainty on the cross-correlation function and are not representative for optimum signal extraction. Indeed, from estimations made by Bouchy et al. (2001) we would expect values which are about 30% lower, i.e. 0.7 m s$^{-1}$ in a 15-minutes integration. This discrepancy finds its origin in the fact that the correlation mask presently used by the data-reduction pipeline is obtained from a G2 star and is not optimized for other spectral types. The measured data and the estimated errors are shown in Table 2.

A periodic variation of the radial velocity was unambiguously detected and characterized during the August 2003 run, and confirmed in February 2004. Amplitude, period and phase of the two data sets match perfectly. The best-fit Keplerian orbit to the data is presented in Fig. 2. It yields a precisely-determined orbital period $P$ of $3.38773 \pm 0.00008$ days and a semi-amplitude of $K = 107 \pm 0.7$ m s$^{-1}$. Using a standard $1/\sigma^2$ weigthening we obtain an *rms* of the data with regard to the Keplerian fit of 2.0 m s$^{-1}$, outlining the high data quality. The complete set of orbital elements with their uncertainties is given in Table 3. The fit results are shown for an eccentricity fixed to zero since releasing the eccentricity in the fit always led to very low and non-significant values, and in particular it did not reduce the residuals at all.

In the fit we allowed for an offset between the mean velocity of the two data sets. The reason for that was that between the two observing runs significant changes to the instrument setup had been made and that we could not exclude a priori the presence of a radial-velocity offset between the data of the two runs. The first GTO run in August took actually place *before* the official start of HARPS (October 1st, 2003). We can report on at least three major instrument setup changes made *after* this run, in September 2003: First, the optical fibers were replaced. Second, the temperature set-point of the spectrograph was decreased from 18 to 17 degrees Celsius. Third, the ThAr calibration lamp had to be replaced by a new one. All of the three operations have been judged critical in the past for an instrument operated in simultaneous Thorium reference mode. Despite the fact that the spectrograph "suffered" these very critical modifications the radial-velocity measurement has not been influenced significantly. The fitted offset is 2.5 m s$^{-1}$. Without this additional free parameter the weighted *rms* on the residuals would have been 2.3 m s$^{-1}$, which is only slightly higher than the final result.

---

[4] The discovery was announced in December 2003 in The ESO Messenger (Mayor et al. 2003), see also www.eso.org/gen-fac/pubs/messenger/archive/no.114-dec03/messenger-no.114.pdf

**Table 2.** HARPS radial-velocity data of HD 330075. The photon-noise error has been estimated on the cross-correlation function of the stellar spectra. The obtained photon noise error is slighlty higher than the expected fundamental limit (Bouchy et al. 2001) mainly because the cross-correlation mask used by the data-reduction pipeline is optimized for a G2 star and not for K1 spectral type.

| Julian Date | RV [km s$^{-1}$] | photon error [km s$^{-1}$] |
|---|---|---|
| 2452851.4874 | 61.3766 | 0.0011 |
| 2452853.4807 | 61.1736 | 0.0029 |
| 2452854.5303 | 61.3297 | 0.0013 |
| 2452854.6556 | 61.3526 | 0.0023 |
| 2452855.6167 | 61.3402 | 0.0033 |
| 2452856.5183 | 61.1850 | 0.0033 |
| 2452856.6590 | 61.1867 | 0.0043 |
| 2452857.4772 | 61.2494 | 0.0017 |
| 2452857.6565 | 61.2875 | 0.0031 |
| 2452858.4613 | 61.3925 | 0.0022 |
| 2452858.6423 | 61.3821 | 0.0021 |
| 2452859.4782 | 61.2495 | 0.0017 |
| 2452859.5022 | 61.2477 | 0.0016 |
| 2452859.6235 | 61.2251 | 0.0018 |
| 2453047.8845 | 61.3632 | 0.0029 |
| 2453048.8853 | 61.3091 | 0.0019 |
| 2453049.8848 | 61.1699 | 0.0018 |
| 2453050.8724 | 61.3048 | 0.0010 |
| 2453052.8718 | 61.2004 | 0.0009 |
| 2453053.8716 | 61.2259 | 0.0010 |
| 2453055.8885 | 61.2641 | 0.0009 |

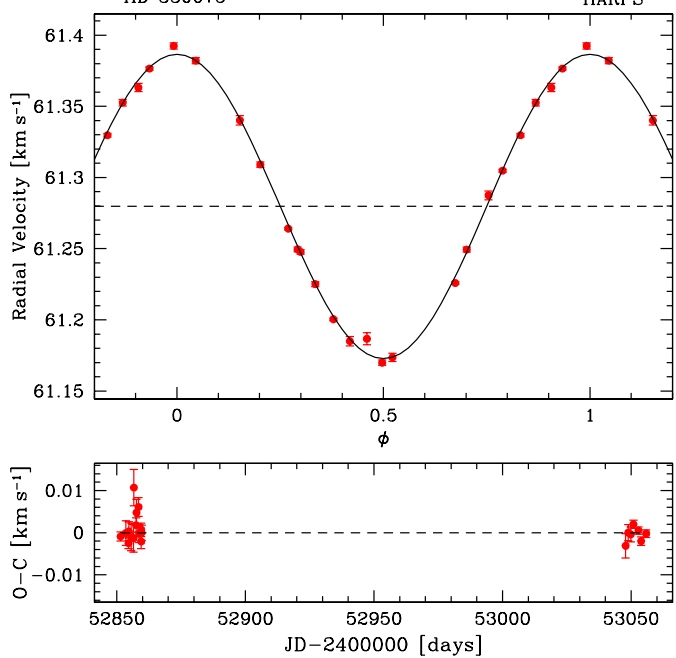

**Fig. 2.** Phase-folded radial-velocity measurements obtained with HARPS for HD 330075. The error bars represent photon-noise errors only, as expected for optimum radial-velocity extraction, and might therefore be underestimated. On the lower panel the residuals of the measured radial velocities to the fitted orbit are plotted as a function of time. In both plots we corrected for an offset between the two observing runs

**Table 3.** HARPS best Keplerian orbital solutions derived for HD 330075 b, as well as inferred planetary parameters

| Parameter | | HD 330075 b |
|---|---|---|
| $P$ | [days] | 3.38773 ± 0.00008 |
| $T$ | [JD] | 2452878.815 ± 0.003 |
| $\gamma$ | [km s$^{-1}$] | 61.2836 ± 0.0004 |
| offset between runs | [m s$^{-1}$] | 2.5 ± 0.8 |
| $e$ | | 0.0 ± fixed |
| $\omega$ | [deg] | 0.0 ± fixed |
| $K$ | [m s$^{-1}$] | 107.0 ± 0.7 |
| $N_{\rm meas}$ | | 21 |
| $\sigma(O\text{-}C)$ | [m s$^{-1}$] | 2.0 |
| $a_1 \sin i$ | [Mm] | 4.983 |
| $f(m)$ | [M$_\odot$] | 4.297·10$^{-8}$ |
| $m_1$ | [M$_\odot$] | 0.7 |
| $m_2 \sin i$ | [M$_{\rm Jup}$] | 0.62 |
| $a$ | [AU] | 0.039 |
| $T_{\rm eq}$ | [K] | 990 |

The reduced $\chi = 1.44$ indicates that we are very close to the photon noise limit. The remaining noise has to also account for possible guiding errors or other external errors. Other error sources could be the calibration procedures and the data reduction, which are still being optimized on the basis of a complete data set we are acquiring during the first year of observations. Finally, we should not forget the star itself: Although not active, the star might contribute to increased residuals, for examples through p-mode oscillations which can attain amplitudes of 1-2 m s$^{-1}$ for a K1 dwarf (Mayor et al. 2003), and even higher amplitudes if the star is evolved.

### 3.3. Planetary characteristics

Using the best-fit orbital parameters and the mass of HD 330075 we derive for the companion a *minimum* mass $m_2 \sin i = 0.62$ M$_{\rm Jup}$. Because of the large time span covered and the low internal error of the data points the orbit is determined very accurately. This allows us to compute the minimum mass of the companion with an accuracy of better than 0.44%, if we neglect the uncertainty on the mass of the primary, which is the major error source. From the orbital parameters and the star mass we get also the separation of the companion to its parent star $a = 0.039$ AU. The surface equilibrium temperature of the planet at such a distance is estimated to be about 990 K, following Guillot et al. (1996).

HD 330075 b belongs to the so-called "hot Jupiter" category of extra-solar planets. The close location to its parent star makes the planet a good candidate for a photometric transit search. Photometric measurements have been carried out during the nights JD = 2452861, 2452862 and 2452879 on the Danish 50cm SAT-telescope at the La Silla Observatory. Unfortunately they did not reveal the presence of a transit (Olsen et al. in prep.).

In order to exclude any kind of activity-induced velocity variations we have performed several tests. First, low activity indicators together with a low projected stellar rotational velocity imply that the velocity scatter is low (Santos et al. 2000). In addition, the stellar rotation period of 48 days, as derived from the activity level of the star, is significantly different from the orbital period, such that we can exclude any correlation. Second, we have verified the bisector of the cross-correlation function with the method described in Queloz et al. (2001a). The bisector remains constant within the photon-noise precision, and, in particular, does not show any in-phase variations with the radial velocity. Finally, the star is photometrically quiet (Olsen et al. in prep.). In summary we are able to exclude that the observed radial-velocity variations are produced by the star itself. The planetary explanation seems therefore to be the most likely.

## 4. Conclusions

In this paper we have presented the first extra-solar planet discovered with HARPS. The planet around HD 330075 is a typical "hot Jupiter" which modulates the radial velocity of its parents star by $\pm 107$ m s$^{-1}$. This signal is easily detected with HARPS. The new instrument especially impresses by its high detection efficiency and the low residuals on the measured orbit. The residuals are as low as 2 m s$^{-1}$, despite the fact that part of the data was obtained inbetween different commissiong phases marked by critical instrument changes. Even better precision is expected on future measurements. Presently, we estimate that HARPS is about 50-100 times more efficient than for instance the CORALIE spectrograph, which has already proved its power during the past years. We expect therefore that HARPS will contribute even more strongly to the quest for extra-solar planets and their characterization during the coming years.

*Acknowledgements.* We are grateful to all technical and scientific collaborators of the HARPS Consortium, ESO Head Quarter and ESO La Silla who have contributed with their extra-ordinary passion and valuable work to the success of the HARPS project. N. C. Santos and D. Naef acknowledge Fundação para Ciência e Tecnologia (Portugal) and Swiss National Science Foundation (FNS grant PBGE2-101322) for their respective support. We would like to thank the European RTN "The Origin of Planetary Systems (PLANETS, contract number HPRN-CT-2002-00308) for supporting this project. This research has made use of the SIMBAD database, operated at CDS, Strasbourg, France.

## References

Baranne, A., Queloz, D., Mayor, M., et al. 1996, A&AS, 119, 373
Bouchy, F., Pepe, F., & Queloz, D. 2001, A&A, 374, 733
Butler, R. P., Bedding, T. R., Kjeldsen, H., et al. 2004, ApJ, 600, L75
Donahue, R. A. 1993, Ph.D. Thesis, New Mexico State University
Endl, M., Cochran, W. D., Tull, R. G., & MacQueen, P. J. 2003, AJ, 126, 3099
Endl, M., Kürster, M., Els, S. H. A. P., et al. 2002, A&A, 392, 671
ESA. 1997, The HIPPARCOS and TYCHO catalogue, ESA-SP 1200
Fischer, D. A., Marcy, G. W., Butler, R. P., et al. 2001, ApJ, 551, 1107
Guillot, T., Burrows, A., Hubbard, W. B., Lunine, J. I., & Saumon, D. 1996, ApJ, 459, L35
Hatzes, A. P., Cochran, W. D., Endl, M., et al. 2003, ApJ, 599, 1383
Henry, T. J., Soderblom, D. R., Donahue, R. A., & Baliunas, S. L. 1996, AJ, 111, 439
Mayor, M., Pepe, F., Queloz, D., et al. 2003, The Messenger, 114, 20
Mayor, M., Udry, S., Naef, D., et al. 2004, A&A, 415, 391
Noyes, R. & et al. 1997, in ASP Conf. Ser. 119: Planets Beyond the Solar System and the Next Generation of Space Missions
Noyes, R. W., Hartmann, L. W., Baliunas, S. L., Duncan, D. K., & Vaughan, A. H. 1984, ApJ, 279, 763
Olsen et al. in prep.
Pepe, F., Mayor, M., Rupprecht, G., et al. 2002, The Messenger, 110, 9
Perrier, C., Sivan, J.-P., Naef, D., et al. 2003, A&A, 410, 1039
Queloz, D., Henry, G. W., Sivan, J. P., et al. 2001a, A&A, 379, 279
Queloz, D., Mayor, M., Udry, S., et al. 2001b, The Messenger, 105, 1
Santos, N., Mayor, M., Naef, D., et al. 2000, A&A, 361, 265
Santos, N. C., Israelian, G., & Mayor, M. 2004, A&A, 415, 1153
Santos, N. C., Mayor, M., Naef, D., et al. 2002, A&A, 392, 215
Tinney, C. G., Butler, R. P., Marcy, G. W., et al. 2002, ApJ, 571, 528
Udry, S., Mayor, M., Naef, D., et al. 2000, A&A, 356, 590
Vogt, S. S., Butler, R. P., Marcy, G. W., et al. 2002, ApJ, 568, 352